\documentclass[twocolumn,preprintnumbers,amsmath,amssymb]{revtex4}

\usepackage{graphicx}
\usepackage{dcolumn}
\usepackage{bm}

\begin{document}

\title{Experimental determination of the dipolar field in Mn$_{12}$-acetate}

\author{S. McHugh}
\author{R. Jaafar}
\author{M. P.  Sarachik}
 \email{sarachik@sci.ccny.cuny.edu}
\affiliation{
Department of Physics\\ City College of New York, CUNY\\
New York, New York 10031, USA}

\author{Y. Myasoedov}
\author{H. Shtrikman}
\author{E. Zeldov}
\affiliation{
Department of Condensed Matter Physics\\
The Weizmann Institute of Science\\
Rehovot 76100, Israel}

\author{R. Bagai}
\author{G. Christou}
\affiliation{
Department of Chemistry\\
University of Florida\\
 Gainesville, Florida 32611, USA}

\begin{abstract}
Crystals of the molecular magnet Mn$_{12}$-acetate are known to contain a small fraction of low- symmetry (minor) species with a small anisotropy barrier against spin reversal.  The lower barrier leads to faster magnetic relaxation and lower coercive field.  We exploit the low coercive fields of the minor species to make a direct determination of the dipole field in Mn$_{12}$-ac.  We find that the dipolar field of a fully magnetized crystal is $51.5 \pm 8.5$ mT, consistent with theoretical expectations.
\end{abstract}
                     
\maketitle

Mn$_{12}$-acetate is one of the first synthesized and best studied examples of a molecular magnet.  Each molecule in a crystal of Mn$_{12}$-ac contains a cluster of twelve Mn atoms surrounded by non-magnetic ligands.  The Mn atoms are coupled antiferromagnetically by superexchange via oxygen bridges, forming a ferrimagnetic cluster at low temperatures with a large spin $S = 10$ and a zero-field barrier to relaxation $U \approx 60$ K \cite{Lis, Sessoli}.  Theoretical estimates and experimental observations have indicated that interactions between the molecules are weak and the system can be described by an effective Hamiltonian \cite{Friedman}
\begin{eqnarray}
\mathcal{H} = -DS_z^2 -g\mu_B B_z S_z + \ldots + \mathcal{H}_\perp
\end{eqnarray}
where $D= 0.6$ K, and $\mathcal{H}_\perp$ is a small symmetry-breaking term that allows tunneling across the anisotropy barrier.  The Mn$_{12}$-acetate molecule can thus be modeled by a double-well potential plotted as a function of angle, with $(2S+1)$ discrete energy levels  corresponding to the quantum-mechanical spin projections along the easy, c-axis of the crystal, $(S_z=10, 9, 8,\ldots -9, -10)$.  The rate of magnetic relaxation toward equilibrium decreases with decreasing temperature and becomes sufficiently slow that hysteretic behavior is found below a blocking temperature $T_B \sim 3$ K.  Steps in the hysteresis loop are observed whenever the magnetic field brings energy levels into resonance allowing the electronic spins to tunnel across the barrier and increasing the relaxation rate.   The spacing between tunneling resonances can be deduced from the Hamiltonian as $\Delta (\mu_0H_z) = D/g\mu_B$, assuming $B_z = \mu_0H_z$.  The observation of these equally spaced steps provided the first evidence for macroscopic quantum tunneling of the magnetization \cite{Friedman}.

Recent work has focused on circumstances where the assumption of independent spins breaks down in crystals of molecular magnets.  Thus, ferromagnetic ordering due to magnetic dipole-dipole interactions has been demonstrated experimentally in high-spin molecular magnets by Morello et al. \cite{Morello} in Mn$_6$ by Evangelisti et al.\cite{Evangelisti} in the high-spin molecular magnet Fe$_{17}$.  Based on neutron scattering experiments, Luis et al. \cite{Luis} have claimed that Mn$_{12}$-ac also orders ferromagnetically provided a large transverse field is applied to reduce the magnetic barrier and increase the relaxation rate.  Recent detailed calculations by Garanin and Chudnovsky\cite{Garanin} indicate that dipolar ferromagnetism in a transverse field should indeed be found in Mn$_{12}$-ac below a Curie temperature $T_c \sim 0.8$ K.  A careful determination of the dipolar fields is therefore timely and important.  In this paper we report the results of a direct measurement of the dipole field obtained by exploiting the properties of the fast-relaxing minor species in crystals of Mn$_{12}$-ac.

It is well known that Mn$_{12}$-ac crystals contain a small fraction of low- symmetry species molecules at a level of roughly 5\% \cite{Minors}.  This ``minor" species, an isomer of the ``major" species of Mn$_{12}$-ac, has a reduced energy barrier of $\approx 42$ K at zero field \cite{Gamma0}.  The magnetic relaxation rates, typically determined by AC susceptibility measurements, are fit with an Arrhenius equation
\begin{eqnarray}
\Gamma = \Gamma_0 \mbox{exp}\left[-U(H)/T\right].
\label{Arrhenius}
\end{eqnarray}
For the major species, $\Gamma_0 \approx 4.2 \times 10^7\mbox{ s}^{-1}$ and $\Gamma_0 \approx 4.5 \times 10^9\mbox{ s}^{-1}$ for the fast-relaxing minor species\cite{Gamma0}.  

\begin{figure}[htbp]
\begin{center}
  \includegraphics[width=3.in, height=2.5in]{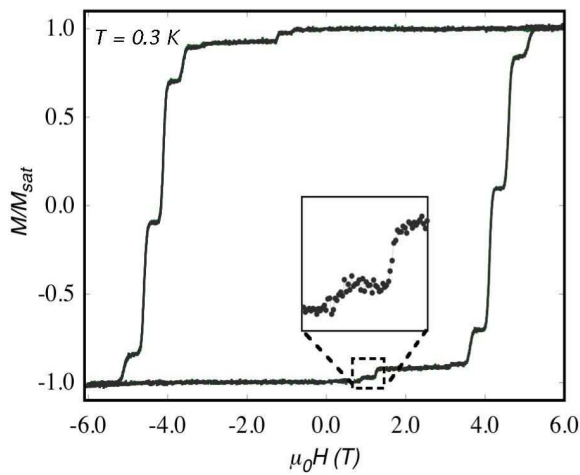}
\caption{(Color online) Magnetization versus magnetic field of a single crystal of Mn$_{12}$-ac measured for an external field sweep rate of $\pm 10$ mT/s.}
\label{fig0}
\end{center}
\end{figure}

\begin{figure}[htbp]
\begin{center}
  \includegraphics[width=3.4in, height=1.75in]{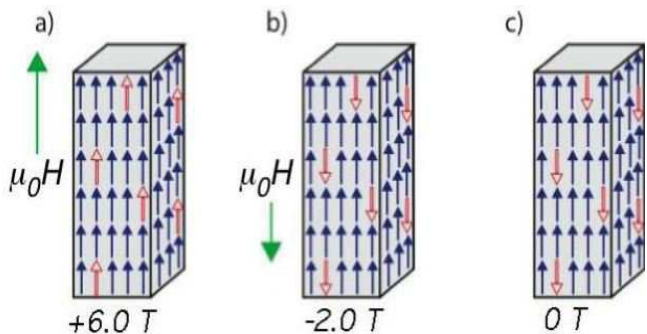}
\caption{(Color online) Schematic of the procedure used to prepare Mn$_{12}$-ac with minor and major species magnetized in opposite directions.  (a) First, a $+6$ T field is applied to align all spins.  (b) Then, a $-2$ T field is applied, reversing the magnetization of the minor species.  (c) Finally, the field is returned to zero, leaving the major and minor species spins anti-parrallel.}
\label{fig1}
\end{center}
\end{figure}

\begin{figure}[htbp]
\begin{center}
  \includegraphics[width=3in, height=5in]{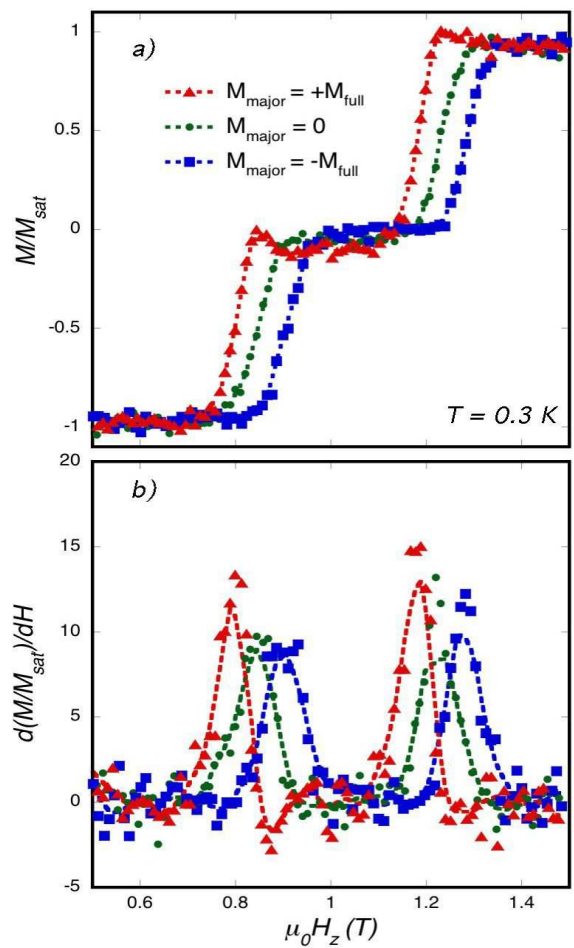}
\caption{(Color online) (a) Magnetization of the minor species as a function of external magnetic field swept at $+5$ mT/s with the major species magnetization prepared following the three protocols described in the text.  The triangles (squares) are data taken with the major species aligned in the positive (negative) direction.  The circles are data taken with the minor species randomly oriented to yield zero magnetization.  (b)  The derivative of the curves shown in frame (a).}
\label{Magnetization}
\end{center}
\end{figure}

The different relaxation rates are also evident in magnetization curves.  Figure \ref{fig0} shows a hysteresis loop for a single crystal of Mn$_{12}$-ac taken at $0.3$ K with an external field sweep rate of $\pm 10$ mT/s.  Starting from zero field, the total magnetization is constant as the field is increased to $0.90$ T.  Small steps are observed at $\sim 0.9$ T and $1.28$ T (see also Fig. 3) corresponding to the resonant relaxation of the minor species magnetization.  Above $1.5$ T, all minor species molecules have reversed and no further change in the total magnetization is seen until $\approx 3$ T.  Between $3.1$ and $5.3$ T, evenly space steps are observed due to tunneling of the major species.  
Above $5.3$ T, all the major species molecules have reversed and the crystal is completely magnetized.  Using a simple protocol similar  to one described in Ref. \cite{Wernsdorfer}, the difference in the fields required to flip the spins of the major and minor species allows one to magnetically prepare the crystal with: (A) minor species aligned and major species spins random; (B) major and minor species spins aligned in the same direction;  and (C) major and minor species spins aligned in opposite directions.

(A) Starting with an unmagnetized (zero-field-cooled) crystal, and maintaining the temperature at $300$ mK:  a magnetic field of $-2$ T is applied which is sufficient to completely magnetize the minor species while leaving the major species spins unchanged; sweeping the field back to zero yields a sample with minor species spins aligned and major species spins random with zero net magnetization.  (B) A larger magnetic field of $-6.0$ T is applied along the c-axis of the crystal, sufficient to magnetize both species; reducing the magnetic field to zero yields a sample with major species and minor species spins aligned in the same direction.  (C) This magnetic preparation is described schematically in Figure \ref{fig1}.  Starting with the spins aligned in same direction (Fig. \ref{fig1}(a)), the magnetic field is ramped to $-2$ T in the direction opposite to the magnetization, reversing the direction of the minor species spins while leaving the major species spins unchanged, as shown in (Fig. \ref{fig1}(b)).  Reducing the field back to zero, (Fig. \ref{fig1}(c)), yields a sample in which the minor and major species are magnetized in opposite directions.

With the sample prepared according to these three methods, we measured the hysteresis loop of the minor species by sweeping the magnetic field between $+2$ and $-2$ T, taking care not to reach fields large enough to reverse the spins of the major species.   The magnetization was measured using an array of micron-sized Hall sensors for single crystal of Mn$_{12}$-ac with typical dimensions of $1.0\times0.3\times0.3$ mm$^3$ immersed in $^3$He.  Experimental details can be found in Ref.\cite{measurements}.

Fig. \ref{Magnetization} (a) shows hysteresis curves for the minor species taken at $0.3$ K and an external field sweep rate of $5$ mT/s for the three different magnetic preparations.  The curves are normalized by $M_{sat}$, the saturation value of the minor species.   The minor species hyteresis curves exhibit a similar staircase structure as the major species.  However, the smaller anisotropy leads to a smaller spacing between steps.  Fig. \ref{Magnetization} (b) shows the derivative of the magnetization curve, which is useful for determining the location and widths of the tunneling resonances.

It is clear that the externally applied magnetic field corresponding to the tunneling resonances depends on the direction of the magnetization of the major species.  The circles of Fig. \ref{Magnetization} are data taken with the net magnetization of the major species equal to zero.  Fully magnetizing the major species in the positive direction shifts the location of the tunneling resonance fields by $\approx -0.05$ T, as indicated by the triangles.   Here the dipolar field adds to the external field to satisfy the resonance condition.  The squares are data taken with the major species magnetized fully in the negative direction with a consequent shift in the resonance field of $\approx +0.05$ T. In this case, the external field is larger to offset the dipolar field in the opposite direction.  A close determination of the shift corresponding to the dipolar field associated with full magnetization of the major species of Mn$_{12}$-ac yields $51.5 \pm8.5$ mT.   The dipolar fields of a completely magnetized crystal of Mn$_{12}$-ac calculated in Ref. \cite{Garanin} was found to be $52.6$ mT, in excellent agreement with our data.  Further evidence is provided by measurements of magnetic avalanches in Mn$_{12}$-ac and will be published elsewhere.

This work was supported at City College by NSF grant DMR-00451605.  E. Z. acknowledges
the support of the Israel Ministry of Science, Culture and Sports.  Support for G. C. was provided by NSF grant CHE-0414555.

\end{document}